\documentclass[prl,aps,twocolumn,floatfix]{revtex4}
\usepackage{amsmath,graphics,epsfig,color,verbatim,ulem}

 \newcommand{\vk}{{\mathbf{k}}}

\begin{document}

\title{Cluster Dynamical Mean Field Theory of the Mott Transition}

\author{H. Park, K. Haule and G. Kotliar}

\affiliation{Department of Physics, Rutgers University,
Piscataway, NJ 08854, USA}

\begin{abstract}
We address the nature of the Mott transition in the Hubbard model
at half-filling using cluster Dynamical Mean Field Theory (DMFT).  We compare
cluster DMFT results with those of single site DMFT.
We show that inclusion of the short range correlations on top of
the on-site correlations, already treated exactly in single site DMFT,
do not change the order of the transition between the paramagnetic
metal and the paramagnetic Mott insulator, which remains first order.
However, the short range correlations reduce substantially the
critical $U$ and modify the shape of the transition lines.
Moreover, they lead to very different physical properties of the
metallic and insulating phases near the transition, in particular in
the region of the phase diagram where the two solutions coexist.
Approaching the transition from the metallic side, we find an
anomalous metallic state with very low coherence scale
at temperatures as low
as $T=0.01t$. The insulating state is characterized by the relatively
narrow Mott gap with pronounced peaks at the gap edge.
\end{abstract}
 \date{\today} \maketitle

The correlation driven metal insulator transition is one of the most
fundamental problems in condensed matter physics, and continues to
receive intensive attention. It is realized in numerous transition
metal oxides and some organic salts, by application of the pressure or
isovalent chemical substitutions~\cite{Imada:98}.  The
metallic state far from the transition is well described by the Fermi
liquid theory, illustrating the wave-like properties of electrons in
solids.  In the insulating side, the electron behaves as a localized
particle.  Near the transition, the effective Coulomb repulsion between
the carriers is of the same order as the kinetic energy term in the
Hamiltonian. This regime probes the dual character of electron, namely
the particle- and wave-like character,
and requires a non-perturbative method for its description.

The nature of the metal to
insulator transition depends strongly on the degree of magnetic
frustration. In the limit of very large magnetic
frustration, the insulating state is a simple paramagnetic state with
local moments carrying $\log(2)$ entropy. The metallic state is a Fermi
liquid with a very heavy mass. The mass increases as the transition is
approached to match the large entropy of the frustrated paramagnetic
insulator. This is the essence of the Brinkman-Rice theory of the
metal insulator transition, which has been
substantially extended by the single site DMFT of the Hubbard model in
the paramagnetic phase~\cite{Georges:96}.
The key predictions of this approach, such as
the existence of a first order line ending in a second order Ising
point, and numerous high temperature crossovers, have been verified
experimentally~\cite{Limelette:2003}.
The first order phase transition in a strongly frustrated situation
has been confirmed by cluster DMFT studies~\cite{Ohashi:07,Parcollet:04}
and by other techniques~\cite{Onoda:03}.

The completely unfrustrated case is also well understood along the
lines first drawn by Slater, and realized in the half filled one band
Hubbard model with only nearest neighbor hoppings.  Here, the metal
insulator transition is driven by the long range magnetic
ordering. The system is insulating and magnetic for arbitrarily
small values of U, as a reflection of the perfect nesting of the band
structure. The insulating gap results from the formation of a spin
density wave that Bragg scatters the electronic quasiparticles.

The character of the metal insulator transition with an
intermediate degree of frustration (when the long range magnetic order
is fully suppressed, but with strong short range magnetic correlations)
remains an open problem.
Qualitative modifications of the character of the transition are
expected, since at low temperatures the paramagnetic insulating state
has very low entropy.
This problem can be addressed by a sharp mathematical formulation studying
the \textit{paramagnetic solution} of the cluster DMFT equations of the
Hubbard model, keeping the short range correlations only. The early cluster DMFT
studies received conflicting answers depending on different cluster
schemes and different impurity solvers
~\cite{Moukouri:01,Zhang:07}.
However, by going to very low temperatures using new algorithmic developments,
we completely settle this question.

\textit{Method:}
To study this problem, we apply cellular dynamical mean field theory
(CDMFT)~\cite{Kotliar:01,Maier:05} to the two-dimensional Hubbard
model, using plaquette as a reference frame.  In this formalism, the
lattice problem is divided into $2\times2$ plaquettes and the lattice
problem is mapped to an auxiliary cluster quantum impurity problem
embedded in a self-consistent electronic bath. The latter is
represented by an $8\times 8$ matrix of impurity hybridization
$\Delta$, which is determined by the condition
\begin{eqnarray}
\Delta(i\omega) = i\omega+\mu-\Sigma_{c}(i\omega)-\left [ \sum_{\tilde{k}} \frac{1}{i\omega + \mu - t_{c}(\tilde{k})
- \Sigma_{c}(i\omega)}\right ] ^{-1}
\label{eqn:scc}
\end{eqnarray}
where $\Sigma_{c}$ is the matrix of cluster self-energies,
$t_c(\tilde{k})$ is the matrix of tight-binding hoppings expressed in
terms of the large unit cell ($2\times 2$) of the cluster, and
$\tilde{k}$ runs over the reduced Brillouin zone of the problem. We
choose the two dimensional square lattice with only the nearest
neighbor hopping $t$.

The cellular DMFT approach has already given numerous insights into
frustrated models of kappa organics~\cite{Kyung:06,Ohashi:07} as well
as the doping driven Mott transition in the Hubbard model, when
treated with a variety of impurity solvers~\cite{Haule_long:2007}. In
this letter, the auxiliary cluster problem is solved with the
numerically exact continuous time quantum Monte Carlo (CTQMC)
method~\cite{Haule:07,Werner:06}.

\begin{figure}[!ht]
\centering{
  \includegraphics[width=0.7\linewidth,clip=]{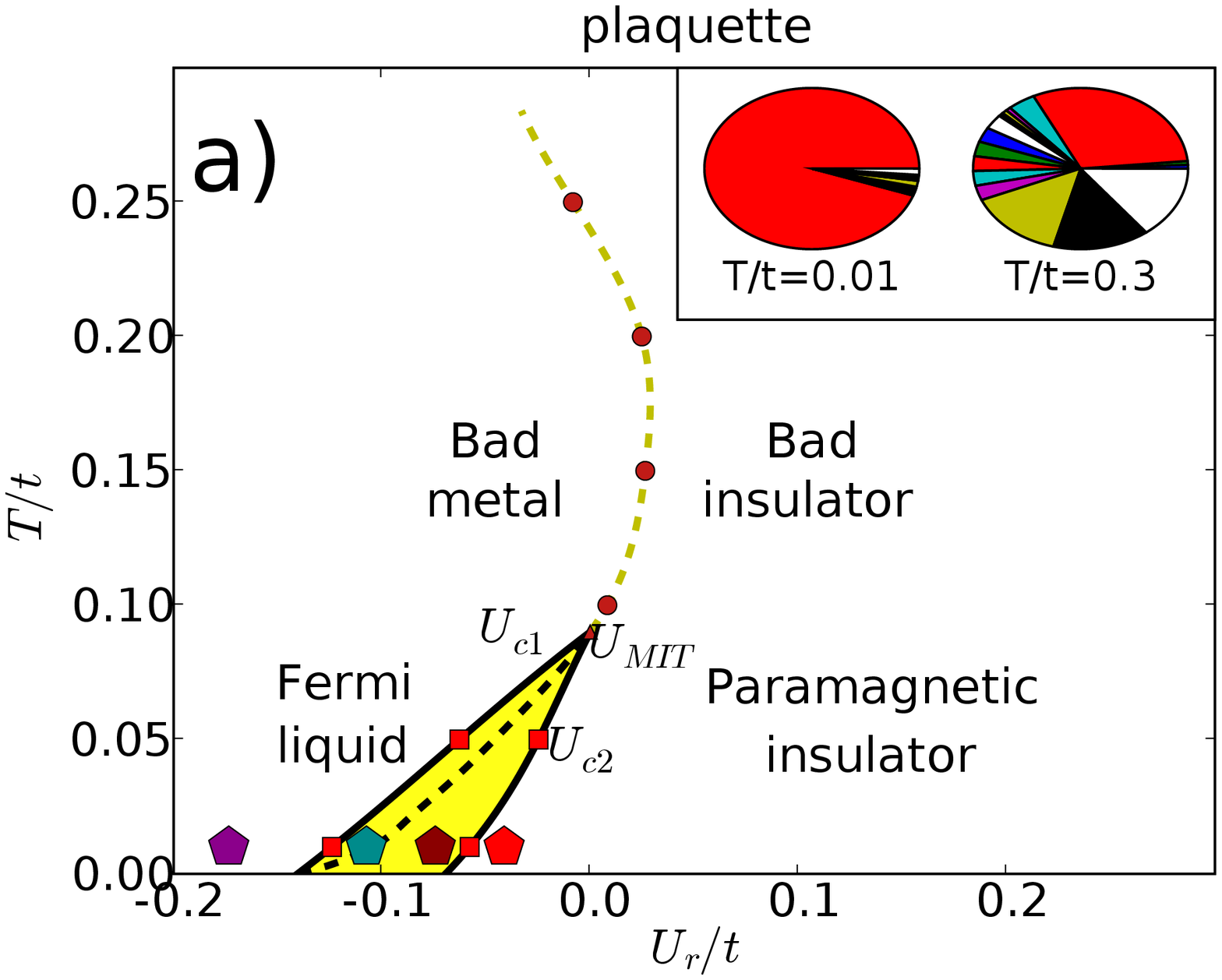}
  \includegraphics[width=0.7\linewidth,clip=]{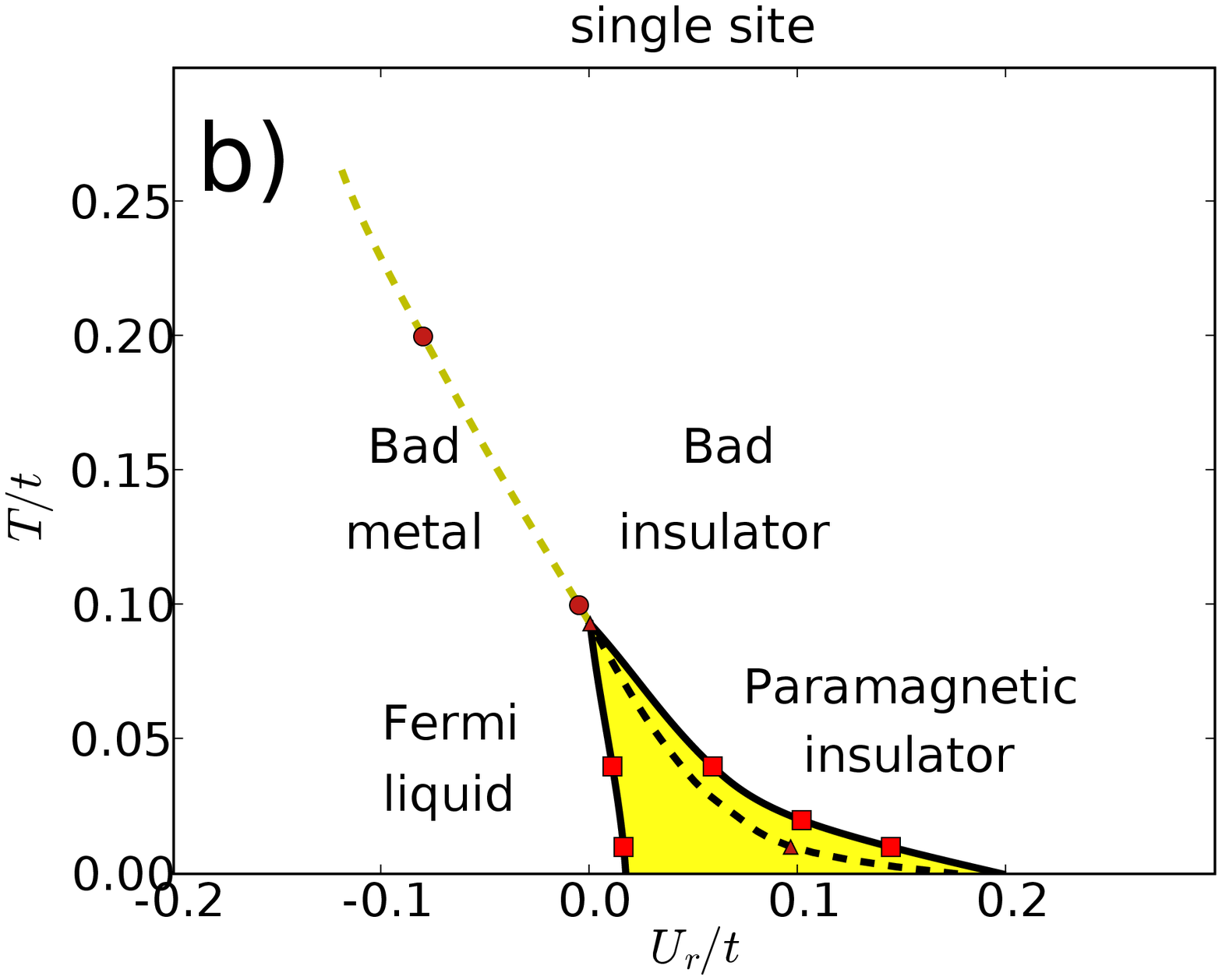}
  }
  \caption{ (Color online) a) The phase diagram of the paramagnetic
    half-filled Hubbard model within plaquette-CDMFT.
    Inset:
      The histogram of the two insulating states. It shows the
      probability for a given cluster eigenstate among the 16
      eigenstates of the half-filled plaquette.  The singlet plaquette
      ground state has the highest probability.
    b) For comparison, the corresponding phase diagram
    of the single site DMFT (using the same 2D density of states)
    is shown. The coexistence region
    is shown as the shaded region.  The dashed line marks the
    crossover above the critical point.  The crossover line was
    determined by the condition that the imaginary part of the
    self-energy at few lowest Matsubara frequencies is flat at the
    crossover value of $U$.  For easier comparison, the x-axis is
    rescaled and the reduced value of
    $U_{r}=\frac{U-U_{MIT}}{U_{MIT}}$ is used. The critical value of
    $U$ is $U_{MIT}=6.05t$ in cluster case and $U_{MIT}=9.35t$ in
    single site case. Pentagons in panel a) mark the points in phase
    diagram for which we present the local spectral functions in
    Fig.2.  }
\label{phased}
\end{figure}

\textit{Results:}
Fig.\ref{phased}a shows the phase diagram of the Hubbard model
within cluster DMFT at half-filling in the absence of long range order.
For interaction strength $U<U_{c2}(T)$, we find a metallic solution
while for $U>U_{c1}(T)$, a Mott insulating solution exists.  The two
transition lines $U_{c1}(T)$ and $U_{c2}(T)$ cross at a second order
endpoint, at temperature $T_{MIT}\sim 0.09t$ and interaction strength
$U_{MIT}\sim 6.05t$.
It is clear that one of the most salient features of the single site
DMFT phase diagram (shown in Fig.~\ref{phased}b), namely the
existence of a first order phase transition, survives in
plaquette-DMFT.

Still there are substantial modifications to the single site DMFT
results when $U/t$ is close to its critical value. Namely,
i) Strong short ranged antiferromagnetic correlations significantly
reduce the value of critical $U$ at which the second order endpoint
occurs.
Note that the plaquette-DMFT critical $U$($\sim 6.05t$) is in very favorable
agreement with the Monte-Carlo crossover $U$ at which the pseudogap
develops at intermediate temperatures accessible by determinantal Monte
Carlo (figure 5 in Ref.~\onlinecite{Vekic:1992}).
This critical $U$ will increase if the system is more frustrated at
short distance. For example, the inclusion of the next nearest hopping
$t'$ has this effect and was studied in
Ref.~\onlinecite{Nevidomskyy:07}.
ii) The shape of the coexistence region, where both metallic and
insulating solutions exist, is significantly different.
The high temperature crossover lines (dashed line above $T\sim 0.1t$
in Fig.\ref{phased}) are similar since at high temperature the entropy of
the paramagnetic insulator is of the order of $\log(2)$ in both cluster
and single site approach.
As the temperature is increased, the large entropy insulating state
wins over the lower entropy metallic state.
At low temperature, the situation is very different.  In single site
DMFT, the metal wins at low temperature in the transition region
because the emergence of the itinerant quasiparticle inside the Mott
gap lowers the free energy of the strongly disordered Mott state.
In the cluster case, the Mott insulator at very low temperature is
very different and has small entropy due to short range singlet
formation. The
small entropy of this state can be confirmed by the "valence histogram"
shown in the inset of Fig.1a. The high temperature insulating state,
which has entropy of the order of $\log(2)$,
populates many states of the plaquette with significant probability.
In contrast, there is only one significant eigenvalue of the density
matrix in low temperature, corresponding to the singlet state.
The insulating phase at low temperature has thus very small entropy,
and the bad metal has larger entropy,
hence decreasing temperature favors insulator over metal.
The actual first order line (dashed
line in Fig.~\ref{phased}a inside the coexistence region, where the
free energy of the two phases equals) therefore bends back and
critical $U$ decreases with decreasing temperature.
It is apparent that the zero temperature
transition in cluster-DMFT happens at $U_{c1}$ and not at $U_{c2}$ as
in DMFT.

While the shape of the DMFT phase diagram strongly resembles the phase
diagram of the Cr-doped V$_2$O$_3$, the reentrant shape of the cluster-DMFT
transition resembles more the $\kappa-$organic
diagram~\cite{Kagawa:04} as pointed out in
Ref.~\onlinecite{Ohashi:07}.

\begin{figure}[!ht]
\centering{
  \includegraphics[width=\linewidth,clip=]{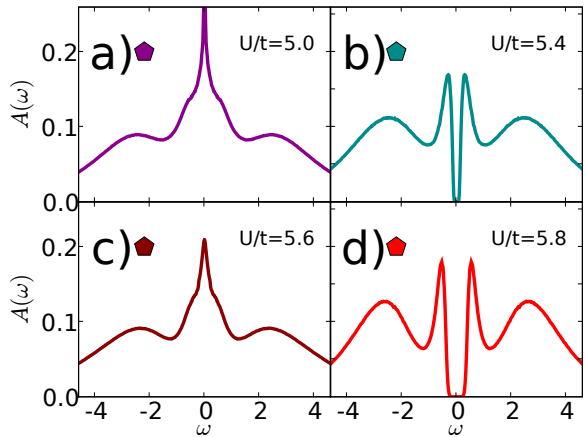}
}
\caption{(Color online) The local spectral function for four
  representative values of $U/t$s and temperature $T=0.01t$ marked by
  pentagons in Fig.1. a) For $U$ below $U_{c1}$ the system is
  in Fermi liquid regime with rather large coherence temperature.
  b) In the coexistence region, the insulating solution
  has a small but finite gap ($\sim 0.2t$). c) The metallic
  solution in the same region is strongly
  incoherent and the value at zero frequency decreases due to
  the finite scattering rate (see self-energy in Fig.~\ref{ImReSig}a).
  d) For $U$ above $U_{c2}$, the Mott gap steadily
  increases with $U$.
  }
\label{Aw}
\end{figure}

To understand the effects brought about by the short range magnetic
correlations near the transition, we focus on the local spectral
functions displayed in Fig.~\ref{Aw}.
As in single site DMFT, below $U_{c1}$ (Fig.~\ref{Aw}a) the system is
a normal Fermi liquid with a reduced width of the quasiparticle peak
($Z \sim 0.4$) and well developed Hubbard bands around $-2.5t$ and
$2.5t$.

The insulator in the coexistence region (Fig.~\ref{Aw}b) is however
very different than Mott insulator in single site DMFT. The Mott gap
is small and it vanishes at $U_{c1}$ where the insulating solution
ceases to exist.
At low temperature very pronounced peaks at the gap edge appear.
These peaks are a clear hallmark of the coherence peaks characteristic
of a Slater spin density wave. This has been noticed earlier in numerous
studies of the Hubbard model~\cite{Hanke:95,Kyung:2006,Zhang:07},
as well as in the single site DMFT
solution in the ordered phase of the unfrustrated lattice, which
captures the physics of perfect nesting.

With increasing $U$ above $U_{c2}$ (Fig.~\ref{Aw}d) the Mott gap
increases but the peaks at the gap edge remain very pronounced. Only
at very large $U$ comparable to the critical $U$ of the single site
DMFT they lose some of their strength and dissolve into a featureless
Hubbard band.

The metallic state, which competes with the insulator in the
coexistence region, (Fig.~\ref{Aw}c) has similar width of the
quasiparticle peak as the Fermi liquid state at $U<U_{c1}$. Hence the
quasiparticle renormalization amplitude, as extracted at finite but
low temperature $T=0.01t$ is rather large. On the other hand, this
metallic solution has somewhat reduced height of the quasiparticle peak
which is mostly due to incoherent nature of the solution.

\begin{figure}[!ht]
\centering{
  \hspace{0.5cm}
  \includegraphics[width=\linewidth,clip=]{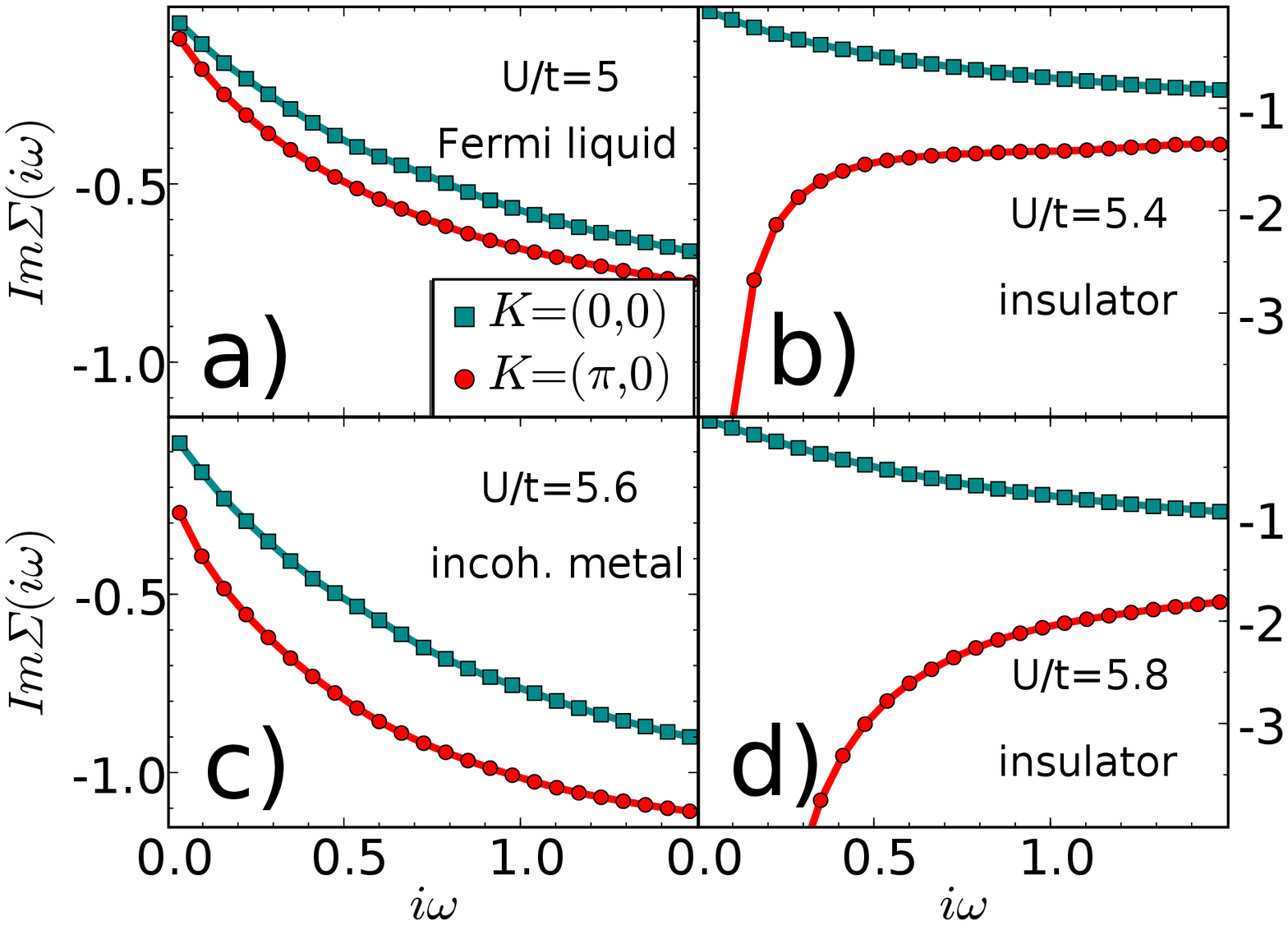}
  \includegraphics[width=0.7\linewidth,clip=]{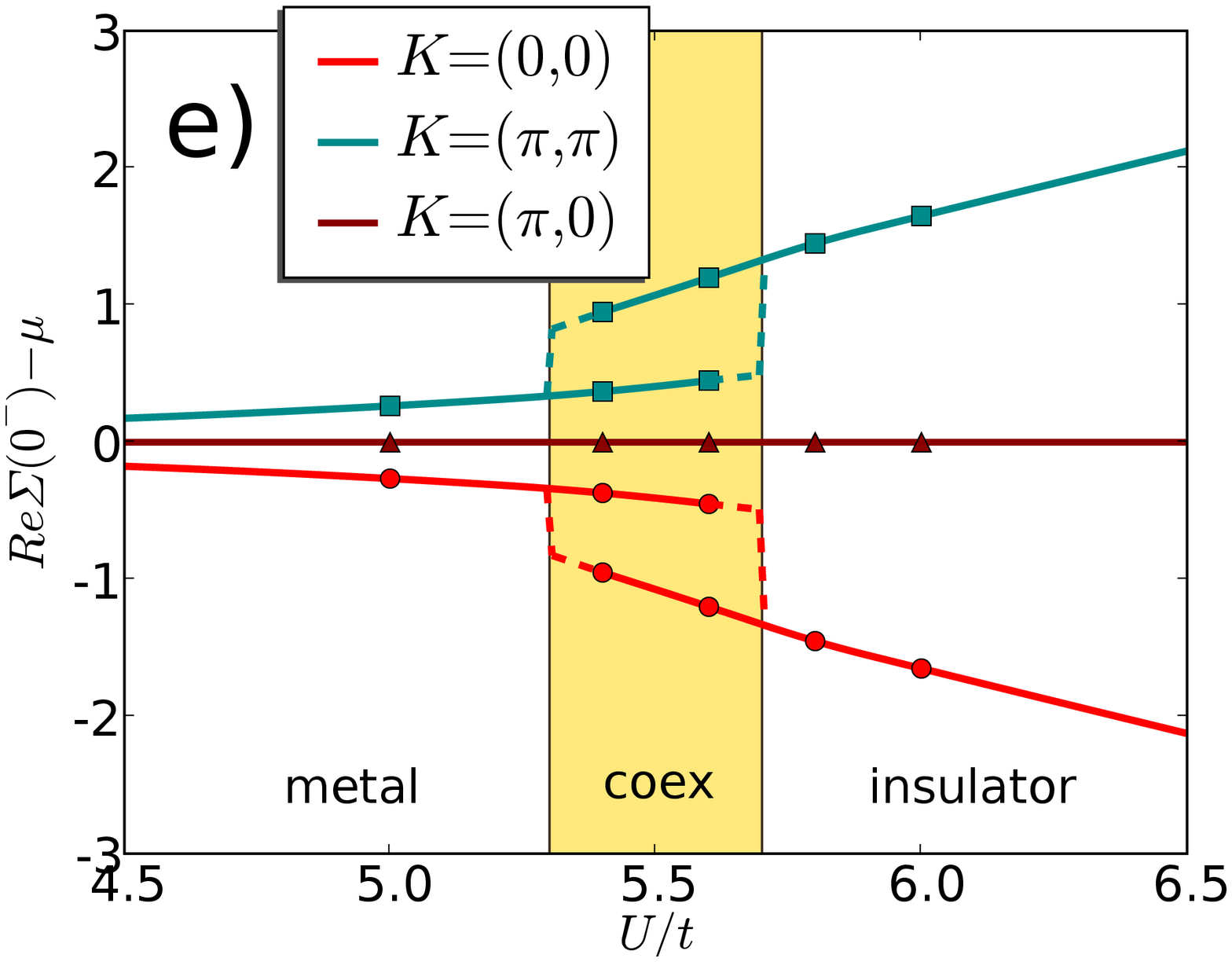}
  }
\caption{(Color online)
 top: The imaginary part of the cluster self
 energies for the same parameters as in Fig.2. Due to particle-hole
 symmetry, the $(\pi,\pi)$ and $(0,0)$ cluster self-energies have the
 same imaginary part and we show only one of them. Below the metal-insulator
 transition shown here in a), the momentum dependence of the self-energy is rather weak
 and the cluster solution is very similar to the single site DMFT
 solution. In the coexistence region, the metallic solution shown here in c)
 is strongly incoherent especially in the $(\pi,0)$ orbital.
 For the insulating solutions in b) and d),
 the $(\pi,0)$ scattering rate diverges which opens the gap in the spectra.
 bottom: e) $Re\Sigma_K(0^{-})-\mu$ as a function of $U$. Due to
 particle-hole symmetry, $Re\Sigma_{K=(\pi,0)}-\mu$ vanishes.
 }
\label{ImReSig}
\end{figure}

The incoherence can also be identified from the raw data on the
imaginary axis. In Fig.~\ref{ImReSig} we show the imaginary
self-energy for the different cluster momenta $K$, which can be
thought as the orbitals of the multi-orbital model associated with the
cluster.
In plaquette geometry, the self-energy is diagonal in
cluster momentum base and the on-site, nearest-neighbor, and
next-nearest-neighbor self-energies can be constructed as the linear
combination of these orbital self-energies\cite{Haule_long:2007}.

Below $U_{c1}$, the self-energies of all four orbitals are very similar
and results are close to the single-site DMFT. The metallic phase in
the coexistence region Fig.~\ref{ImReSig}c has a large scattering rate
in the $(\pi,0)$ orbital, in the orbital which contributes most of the
spectral weight at the fermi level.  The coherence scale in this
strongly incoherent metal is thus severely reduced. The scattering
rate as a function of temperature is not quadratic even at $T=0.01t$
and remains large $\sim 0.2t$ at that temperature.

In Fig.~\ref{ImReSig}b,d the Mott insulating state can be identified by the
diverging imaginary part of the $\Sigma_{(\pi,0)}(i\omega)$. Due to
particle hole symmetry, the real part of the same quantity vanishes.
Therefore, the only way to open a gap in the single particle spectrum
is to develop a pole at zero frequency $\Sigma_{(\pi,0)}\simeq
C/(i\omega)$. We checked that the insulating state in the coexistence
region has the characteristic $1/(i\omega)$ behavior at very low
temperature and
the coefficient $C$ in the coexistence region decreases as $U$
decreases. The closure of the gap at the $U_{c1}$ transition point is
confirmed by the vanishing of $C$ at that point.

The other two orbitals expel their Fermi surfaces by a different
mechanism identified in Ref.~\onlinecite{Haule:2002}, namely the real
parts of the self-energy are such that the effective chemical
potential $\mu_{eff}=\mu-\Sigma(0^-)$ moves out of the band.  The
separation of the two orbitals gradually increases as $U$ increases,
and it jumps at the critical $U$ showing the hysteresis behavior
displayed in Fig.~\ref{ImReSig}e.

\begin{figure}[!ht]
\centering{
  \includegraphics[width=0.7\linewidth,clip=]{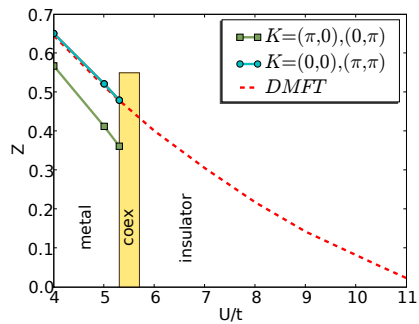}
  }
\caption{ (Color online) The quasi-particle residue $Z$ vs $U/t$ for
  different orbitals in CDMFT. Below the transition point, the $(0,0)$
  and $(\pi,\pi)$ orbitals have essentially the same $Z$ as single
  site DMFT (dotted line) while the quasiparticles are more renormalized in
  $(\pi,0)$ orbital.
  }
\label{Zk}
\end{figure}

The important issue in the metal insulator transition (MIT) is whether
the short range magnetic exchange in the Hubbard type of models allow the
Brinkman-Rice scenario of diverging effective mass.
In Fig.~\ref{Zk} we plot the
quasiparticle renormalization amplitude $Z$ of the four different
orbitals of the plaquette.
As shown in Fig.~\ref{Zk}, the growth of the effective mass in cluster DMFT
is cut-off by the exchange interaction
and the spatial coherence is lost way before the quasiparticles
acquire a large effective mass. The lattice $Z_\vk$ is a linear combination
of the two values plotted in Fig.~\ref{Zk}. The quasiparticles at
$(\pi,0)$ and $(0,\pi)$ are renormalized more strongly than those away from
the two points. More importantly, close to $U_{c1}$, where the system is
still coherent at $T=t/100$, the quasiparticle renormalization
amplitude is rather large for the plaquette without frustration
($Z\sim 0.36$). Very near and inside the coexistence region, the
metallic state remains very incoherent at our lowest temperature
$T=t/100$. We therefore can not determine the low energy $Z$ which
might vanish at $U_{c2}$ at zero temperature.

In conclusion, we used essentially exact numerical method, continuous
time quantum Monte Carlo, and clarified the nature of the Mott
transition in plaquette-DMFT.
The short range correlations which are
accounted for in this study but are absent in single site DMFT do not
change the order of the Mott transition which remains first order
with coexistence of metallic and insulating solution.
Our cluster DMFT
study predicts the existence of anomalous metallic state within the
coexistence region with very low coherence temperature.
This regime could be relevant to the interpretation of experiments in
$V O_2$~\cite{Basov:06} and PrNiO$_{3}$ under the applied
pressure~\cite{Zhou:05} where an anomalous metallic state was
reported.
On the theoretical side, the plaquette DMFT brings new light on the
nature of the interaction driven MIT. The cluster DMFT of this problem
retains aspects of Mott physics, as described in single site DMFT, and
Slater physics. It does that by having two orbitals ($(\pi,0)$ and
$(0,\pi)$) exhibit a Mott transition while the remaining orbitals
($(0,0)$ and $(\pi,\pi)$) undergo a band transition. This Slater-Mott
transition requires momentum space differentiation and has no
analog in single site DMFT.

Acknowledgment: We would like to acknowledge useful discussion with
A.M. Tremblay, V. Dobrosavljevic, and L. de'Medici. G. K. was
supported by NSF grant No. DMR 0528969. H. Park was supported by
Korea Science and Engineering Foundation Grant No. KRF-2005-215-C00050.

\bibliography{plaquette5}

\end{document}